# Quick Annotator: an open-source digital pathology based rapid image annotation tool

Runtian Miao[1], Robert Toth[2], Yu Zhou[1], Anant Madabhushi[1,3], Andrew Janowczyk[1,4]
[1]Case Western Reserve University, Department of Biomedical Engineering, Cleveland OH, USA
[2]Toth Technology LLC, Dover, NJ, USA
[3]Louis Stokes Veterans Administration Medical Center, Cleveland, OH, USA
[4]Lausanne University Hospital, Precision Oncology Center, Vaud, Switzerland

## Abstract

Image based biomarker discovery typically requires an accurate segmentation of histologic structures (e.g., cell nuclei, tubules, epithelial regions) in digital pathology Whole Slide Images (WSI). Unfortunately, annotating each structure of interest is laborious and often intractable even in moderately sized cohorts. Here, we present an open-source tool, Quick Annotator (QA), designed to improve annotation efficiency of histologic structures by orders of magnitude. While the user annotates regions of interest (ROI) via an intuitive web interface, a deep learning (DL) model is concurrently optimized using these annotations and applied to the ROI. The user iteratively reviews DL results to either (a) accept accurately annotated regions, or (b) correct erroneously segmented structures to improve subsequent model suggestions, before transitioning to other ROIs. We demonstrate the effectiveness of QA over comparable manual efforts via three use cases. These include annotating (a) 337,386 nuclei in 5 pancreatic WSIs, (b) 5,692 tubules in 10 colorectal WSIs, and (c) 14,187 regions of epithelium in 10 breast WSIs. Efficiency gains in terms of annotations per second of 102x, 9x, and 39x were respectively witnessed while retaining f-scores >.95, suggesting QA may be a valuable tool for efficiently fully annotating WSIs employed in downstream biomarker studies.

## Introduction

The discovery of biomarkers associated with diagnosis, prognosis, and therapy response from digital pathology whole slide images (WSI) often requires extracting features from precise segmentations of the histologic structures contained within them (e.g., cell nuclei boundaries, tubule shapes, and regions of epithelium)[1–4]. Manually annotating each instance of these histologic structures rapidly becomes intractable, even in small cohorts. For example, the number of nuclei in a single WSI can order into the hundreds of thousands, making accurately individually annotating each cell unfeasible. While a number of image analysis based algorithms have been proposed to help reduce annotation effort, they are not yet integrated into tools with user interfaces enabling their employ[5–8]. Other efforts have resulted in proprietary closed-source tools[9,10] which can be too costly to purchase in academic settings, or do not provide an open environment for facile testing and integration of new algorithms. Other approaches provide command line scripts[11] which are not readily employable by lay users, while additionally requiring the colocation of the user, data, and compute-infrastructure. It is thus clear that an all-inclusive computational tool designed for remote access by pathologists to aid in this annotation process at scale are needed.

Recognizing the need for a modular, user-friendly, annotation tool which significantly accelerates annotation tasks, we present here an open-source image annotation application, Quick Annotator (QA). QA is able to provide significant improvements in annotation efficiency by intelligent usage of deep learning (DL), a form of supervised machine learning which involves multiple neural network layers. QA involves integrating DL[12] with active learning[13], an interactive supervised approach for training machine learning approaches based off selective user feedback. As the user annotates structures in the web-browser based frontend, a popular DL model (u-net[14]) is trained in the backend. This DL model then makes predictions highlighting the structure, allowing the user to either accept or refine pixel-level boundaries in a rapid fashion. This approach allows the DL model to provide feedback to the user, accentuating regions in the image which require additional user input to maximally improve the performance of the next iteration of the supervised classifier.

Through this iterative active learning based process, QA empowers the end user to spend more time efficiently verifying, as opposed to painstakingly annotating histologic structures.

To aid in the annotation process, a number of common image annotation tools are provided, such as brushes and erasers of various sizes, along with polygon style annotation tools. More interestingly, QA also provides the option of highlighting image regions via the selection of DL derived superpixels[15], which are incrementally improved as the DL model improves, facilitating high-fidelity pixel level boundary selection (**Figure S4**). Importantly, QA is designed in an especially modular way, such that as improvements in both DL technology and architectures are discovered, they can rapidly be integrated into QA with minimal modifications to the base application. The output from QA can immediately be used for downstream applications such as feature extraction or training a larger more sophisticated DL segmentation model. In this work we demonstrate the utility of QA for segmentation at three scale lengths typical in computational pathology (**Table 1** and **Figure S1**). At the lowest length scale, from 5 WSI corresponding to pancreatic cancer, 337,386 nuclei were segmented. For the intermediate scale, from 10 WSI containing colorectal cancer, 5,692 tubules were segmented. Lastly for the largest scale, 14,187 regions of epithelium, totaling an area of 35,844,637 pixels, were segmented in 10 WSI images.

| Tissue scale | Histologic structure | Number of slides | Number of ROIs | Number of histologic structures | QA total time (min) | QA human time ($QA_t$, min) | Manual time ($M_t$, min) | Speed up ($\theta_t$) | F-score |
|---|---|---|---|---|---|---|---|---|---|
| **Small** | Cell Nuclei | 5 | 400 | 337,386 | 473 | 391 | 40,165 | 102X | 0.97 |
| **Medium** | Tubules | 10 | 100 | 5,692 | 121 | 101 | 923 | 9X | 0.95 |
| **Large** | Epithelium | 10 | 100 | 14,187 | 167 | 113 | 4,433 | 39X | 0.89 |

**Table 1.** Description of the datasets used for validation of QA along with the demonstrated speedup. As mentioned above, the difference between QA total time and QA human time is that human-time removes DL training time, as the human annotator was dismissed to perform other non-related annotation tasks. On the other hand, QA total time includes model training time under the assumption that the user kept annotating during backend training. Manual time is derived by extrapolating the measured annotations per minute from a subset of the annotations.

## Materials and Methods

In accordance with the QA workflow (Section **SM1**), each WSI image was broken into tiles and processed individually. To begin, tiles originating from the same WSI image were uploaded into QA, which divided these tiles into smaller 256 x 256 patches. A u-net consisting of a block depth of 5 layers and 113,306 parameters was trained on these image patches in an auto-encoding fashion to produce a baseline model, a process shown to learn features associated with tissue presentation[16]. This base model is subsequently fine-tuned in a supervised fashion to segment the structure of interest, a common approach to help reduce annotated data requirements[17]. Next, the user viewed all patches processed by this model in a uniform manifold approximation and projection (UMAP[18]) plot (**Figure 1A**). This plot maps the high dimensional space learned by the DL model into a 2-dimensional representation such that patches perceived to be similar by the model are plotted proximally. The user selects dispersed patches for annotation (**Figure S3**) to improve training set diversity. As the user annotates these patches (**Figure 1B**), the DL model begins to make suggestions which can be accepted or modified (**Figure 1C**).

The time to annotate the tiles is recorded to estimate a metric of structures per second, which is reported both in total time and human time. The difference between these metrics being that human-time removes the DL model training time, as the human annotator can be dismissed to perform other unrelated tasks.

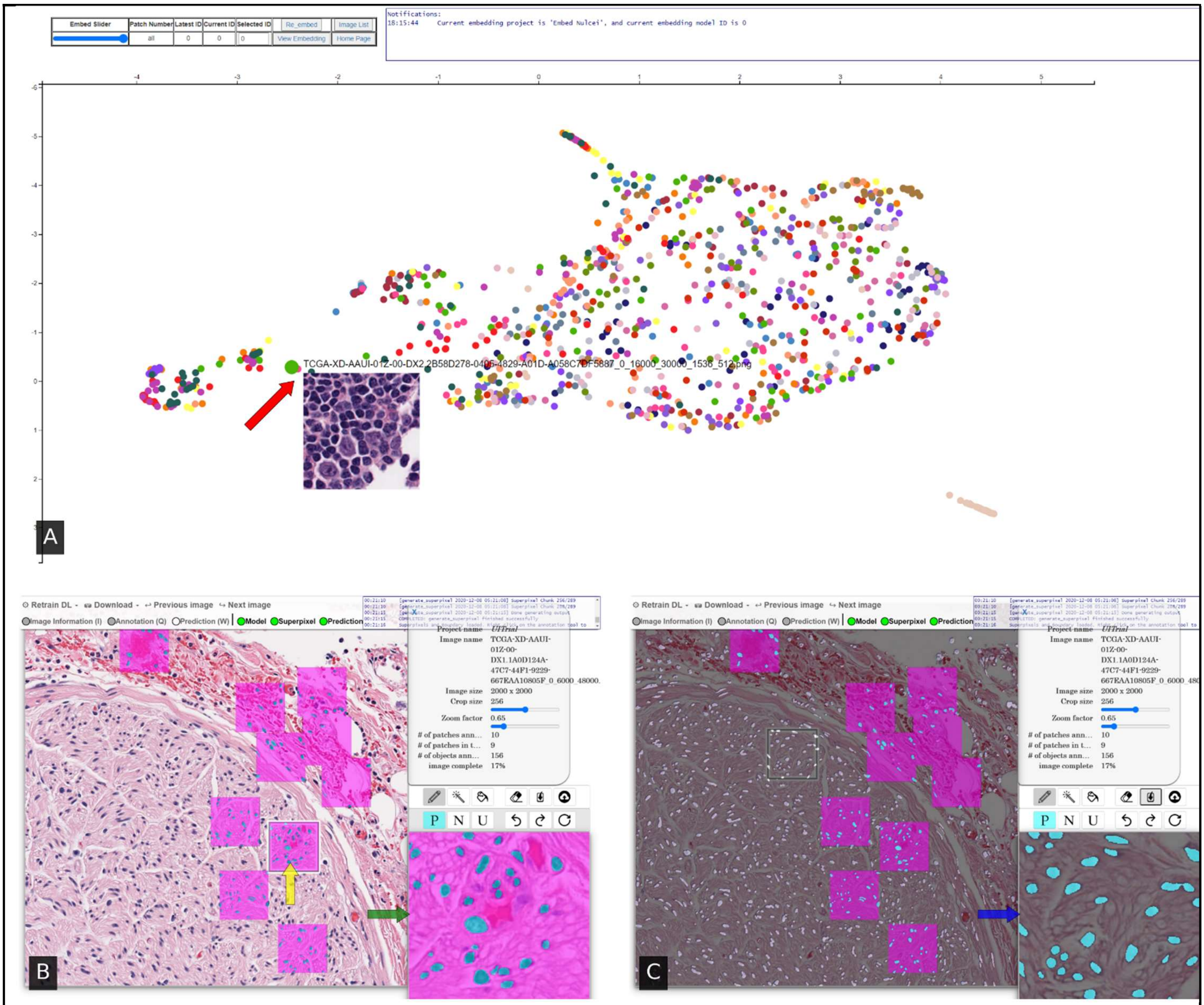

**Figure 1.** The user interface consists of (A) an embedding plot page, and (B & C) an annotation page. The embedding page is a 2D representation of all patches in the system, where patches perceived to be similar by the model are plotted closely together, and updates as the model trains. By actively annotating patches across the space, the user provides more diverse training exemplars to the DL model, thus facilitating the creation of a more robust model sooner. When hovering over a dot, a preview of the patch is shown (red arrow), and subsequently clicking on the dot takes the user to the (B) annotation page centered around that patch. There, the user selects (yellow arrow) a square region which is then loaded into the high magnification annotation window (green arrow). The user annotates positive regions in turquoise and non-target areas in fuchsia using common image markup tools. After annotating at least 3 patches, the user can train a DL classifier to generate annotation suggestions in (C) white overlay. The user may then import the classifier's suggestions into annotation window (blue arrow) and edit, if needed, before accepting.

To form a manual baseline for comparison, an open-source digital pathology tool, QuPath[19], was employed. QuPath is one of the most widely used open-source image analysis toolkits used by researchers and pathologists, owing to its highly polished and intuitive interface, cross platform support, and ease of execution of common analytical workflows. In each use case, QuPath was used to annotate a subset of the data, forming the ground truth for comparison. Additionally, the manual time needed to annotate this subset was recorded and used to compute an approximate total annotation time ($M_t$) needed for completion of the entire task. Quantitatively, efficiency improvement was defined as the ratio ($\theta_t$) between $M_t$ and QA time ($QA_t$). Pixel-level f-scores were reported comparing the masks created in the manually annotated subset of data with that of QA-aided annotations to ensure comparable annotations were produced.

## Results and Discussion

Our results indicate (a) the speed efficiency improvement afforded by QA is significant, and (b) QA annotations remained highly concordant with those produced manually. It is important to note that the user is the final arbiter of what is an acceptable annotation, and always has the ability to manually adjust any pixel that they are in disagreement with. Interestingly, differences still remain in the reported f-scores between manual and QA produced masks. This can be attributed to the higher-level of precision afforded by computational tools (**Figure 2** and **Figure S5**), a phenomenon others have reported[12,20], and which further represents an interesting opportunity for such tools to significantly improve the quality of annotated datasets.

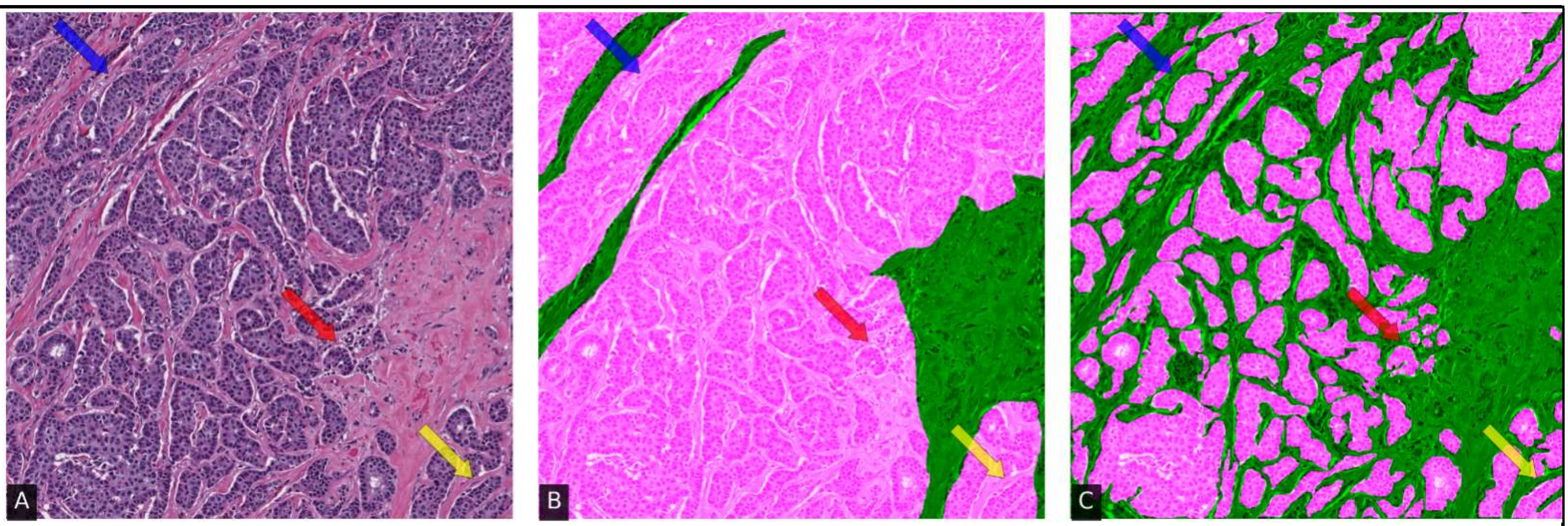

**Figure 2.** The (A) original 1000 x 1000 epithelium ROI with associated (B) manual and (C) QA annotation overlaid in fuchsia, with an f-score of .68. In intricate epithelial regions (e.g., areas indicated with arrows), the QA classifier appears to be able to provide annotation suggestions at a level of precision that would not be tractable for a user to perform manually.

Usage of QA appears to proceed in two distinct workflows. At the beginning, the user is required to provide individual manual annotations, as the model itself is not sufficiently exposed to a representative set of training exemplars. In this workflow, QA and manual annotation efficiency are comparable. This workflow quickly transitions (**Figure 3**) to one wherein the user effort is more focused on reviewing and accepting predictions from the DL model. Here the efficiency gains appear to greatly improve, as with a singular click the user can accept large regions containing many structures, as opposed to manually interacting with each of them. This behavior is seen in all use cases, and interestingly creates a point of discussion on the suitability of tasks for QA. In those cases where very few structures of interest are present, such as delineating a single large tumoral region, there is likely minimal value in employing QA, and instead the slides should be annotated manually. On the other hand, when the number of individual annotations required rises, it becomes evident that employing QA may result in significant efficiency gains.

Worth noting is that post-installation of QA, no internet connection is required, thus making it suitable for non-anonymized clinical data. In fact, given the modestly sized DL networks employed, and its operating system agnostic design, recently purchased laptops are sufficiently powerful to use QA. In spite of this, one can easily host QA on a server with a powerful graphics processing unit (GPU), thus enabling remote access for, e.g., clinical pathologists, to collect annotations (i.e., bringing the expert to the data) without the need for the local download and manipulation of large amounts of data (i.e., bringing the data to the expert), an often burdensome paradigm. A helpful consequence of this approach is that no software needs to be installed locally, which is often heavily restricted in clinical environments.

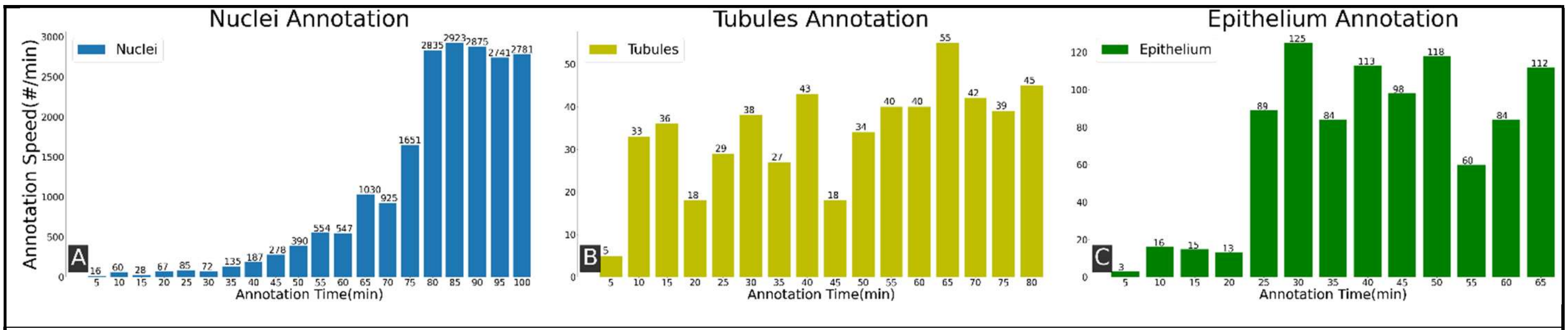

**Figure 3**. Efficiency metric over time demonstrating the improvement in speed afforded by QA in annotating (A) nuclei, (B) tubules, and (C) epithelium. The x-axis is the human annotation time in minutes and the y-axis is the annotation speed in terms of annotated histologic structures per minute. The trend of performance improvements varies per use case with (a) the nuclei showing a consistent improvement in time, (b) the tubule performance plateauing after annotating a few structures, and (c) the epithelium requiring a number of additional iterations before reaching its plateau. These plateaus indicate the DL model is sufficiently trained to produce suggestions agreeable to the user.

In conclusion, Quick Annotator is a high-throughput image annotation tool being publicly released for community review, comment, and usage. QA has demonstrated significant improvements in annotation efficiency, without sacrificing annotation fidelity, and in fact often improves upon what may be possible for humans to complete without computer aided tools. Future versions of QA are aimed at incorporating support for directly annotating WSI, as well as further hiding the latency of DL training from the user perspective. The source code of Quick Annotator is freely available for use, modification, and contribution*.

## References

1. Whitney, J. et al. Quantitative nuclear histomorphometry predicts oncotype DX risk categories for early stage ER+ breast cancer. BMC Cancer 18, 610 (2018).
2. Lu, C. et al. Feature-driven Local Cell Graph (FLocK): New Computational Pathology-based Descriptors for Prognosis of Lung Cancer and HPV Status of Oropharyngeal Cancers. Medical Image Analysis 101903 (2020) doi:10.1016/j.media.2020.101903.
3. Javed, S. et al. Cellular community detection for tissue phenotyping in colorectal cancer histology images. Medical Image Analysis 63, 101696 (2020).
4. Failmezger, H. et al. Topological Tumor Graphs: A Graph-Based Spatial Model to Infer Stromal Recruitment for Immunosuppression in Melanoma Histology. Cancer Res 80, 1199–1209 (2020).
5. Carse, J. & McKenna, S. Active Learning for Patch-Based Digital Pathology Using Convolutional Neural Networks to Reduce Annotation Costs. in Digital Pathology (eds. Reyes-Aldasoro, C. C., Janowczyk, A., Veta, M., Bankhead, P. & Sirinukunwattana, K.) 20–27 (Springer International Publishing, 2019). doi:10.1007/978-3-030-23937-4_3.
6. Das, A., Nair, M. S. & Peter, D. S. Batch Mode Active Learning on the Riemannian Manifold for Automated Scoring of Nuclear Pleomorphism in Breast Cancer. Artif Intell Med 103, 101805 (2020).
7. Pati, P., Foncubierta-Rodríguez, A., Goksel, O. & Gabrani, M. Reducing annotation effort in digital pathology: A Co-Representation learning framework for classification tasks. Medical Image Analysis 67, 101859 (2021).
8. Van Eycke, Y.-R., Foucart, A. & Decaestecker, C. Strategies to Reduce the Expert Supervision Required for Deep Learning-Based Segmentation of Histopathological Images. Front Med (Lausanne) 6, 222 (2019).
9. Lindvall, M. et al. TissueWand, a Rapid Histopathology Annotation Tool. J Pathol Inform 11, 27 (2020).

*The source code is available at https://github.com/choosehappy/QuickAnnotator.

10. Martel, A. L. et al. An Image Analysis Resource for Cancer Research: PIIP -- Pathology Image Informatics Platform for Visualization, Analysis and Management. Cancer Res 77, e83–e86 (2017).
11. Lutnick, B. et al. An integrated iterative annotation technique for easing neural network training in medical image analysis. Nat Mach Intell 1, 112–119 (2019).
12. Janowczyk, A. & Madabhushi, A. Deep learning for digital pathology image analysis: A comprehensive tutorial with selected use cases. J Pathol Inform 7, (2016).
13. Doyle, S., Monaco, J., Feldman, M., Tomaszewski, J. & Madabhushi, A. An active learning based classification strategy for the minority class problem: application to histopathology annotation. BMC Bioinformatics 12, 424 (2011).
14. Ronneberger, O., Fischer, P. & Brox, T. U-Net: Convolutional Networks for Biomedical Image Segmentation. in Medical Image Computing and Computer-Assisted Intervention – MICCAI 2015 (eds. Navab, N., Hornegger, J., Wells, W. M. & Frangi, A. F.) 234–241 (Springer International Publishing, 2015). doi:10.1007/978-3-319-24574-4_28.
15. Achanta, R. et al. SLIC superpixels compared to state-of-the-art superpixel methods. IEEE Trans Pattern Anal Mach Intell 34, 2274–2282 (2012).
16. Janowczyk, A., Basavanhally, A. & Madabhushi, A. Stain Normalization using Sparse AutoEncoders (StaNoSA): Application to digital pathology. Computerized Medical Imaging and Graphics (2016) doi:10.1016/j.compmedimag.2016.05.003.
17. Kandel, I. & Castelli, M. How Deeply to Fine-Tune a Convolutional Neural Network: A Case Study Using a Histopathology Dataset. Applied Sciences 10, 3359 (2020).
18. McInnes, L., Healy, J., Saul, N. & Grossberger, L. UMAP: Uniform Manifold Approximation and Projection. The Journal of Open Source Software 3, 861 (2018).
19. Bankhead, P. et al. QuPath: Open source software for digital pathology image analysis. Scientific Reports 7, 16878 (2017).
20. Doyle, S., Feldman, M., Tomaszewski, J. & Madabhushi, A. A boosted Bayesian multiresolution classifier for prostate cancer detection from digitized needle biopsies. IEEE Trans Biomed Eng 59, 1205–1218 (2012).

## Conflict of Interest Statement

AM is an equity holder in Elucid Bioimaging and in Inspirata Inc. In addition, he has served as a scientific advisory board member for Inspirata Inc, Astrazeneca, Bristol Meyers-Squibb and Merck. Currently he serves on the advisory board of Aiforia Inc. He also has sponsored research agreements with AstraZeneca, Bristol Meyers-Squibb and Boehringer-Ingelheim. His technology has been licensed to Elucid Bioimaging. He is also involved in a NIH U24 grant with PathCore Inc, and 3 different R01 grants with Inspirata Inc. AJ provides consulting services for Merck and Roche. No other conflicts of interest were declared. The content is solely the responsibility of the authors and does not necessarily represent the official views of the National Institutes of Health, the U.S. Department of Veterans Affairs, the Department of Defense, or the United States Government.

## Acknowledgements

Research reported in this publication was supported by the National Cancer Institute under award numbers: 1U24CA199374-01, R01CA202752-01A1, R01CA208236-01A1, R01 CA216579-01A1, R01 CA220581-01A1, 1U01 CA239055-01, 1U01CA248226-01, 1U54CA254566-01the Ohio Third Frontier Technology Validation Fund
National Heart, Lung and Blood Institute 1R01HL15127701A1
National Institute for Biomedical Imaging and Bioengineering 1R43EB028736-01
National Center for Research Resources under award number 1 C06 RR12463-01
VA Merit Review Award IBX004121A from the United States Department of Veterans Affairs Biomedical Laboratory Research and Development Service
the Office of the Assistant Secretary of Defense for Health Affairs, through
the Breast Cancer Research Program (W81XWH-19-1-0668)
the Prostate Cancer Research Program (W81XWH-15-1-0558, W81XWH-20-1-0851)
the Lung Cancer Research Program (W81XWH-18-1-0440, W81XWH-20-1-0595)
the Peer Reviewed Cancer Research Program (W81XWH-18-1-0404)

the Kidney Precision Medicine Project (KPMP) Glue Grant
the Ohio Third Frontier Technology Validation Fund
the Clinical and Translational Science Collaborative of Cleveland (UL1TR0002548) from the National Center for Advancing Translational Sciences (NCATS) component of the National Institutes of Health and NIH roadmap for Medical Research
The Wallace H. Coulter Foundation Program in the Department of Biomedical Engineering at Case Western Reserve University.
The authors also thank the NVIDIA Corporation for the gift of two Titan-X GPUs.
The content is solely the responsibility of the authors and does not necessarily represent the official views of the National Institutes of Health, the U.S. Department of Veterans Affairs, the Department of Defense, or the US Government.

## Supplementary Material

### SM 1. Experiment setup and workflow:

In this paper we focused on 3 histologic structures for segmentation: pancreatic nuclei, colorectal tubules, and breast cancer (see **Figure S1**).

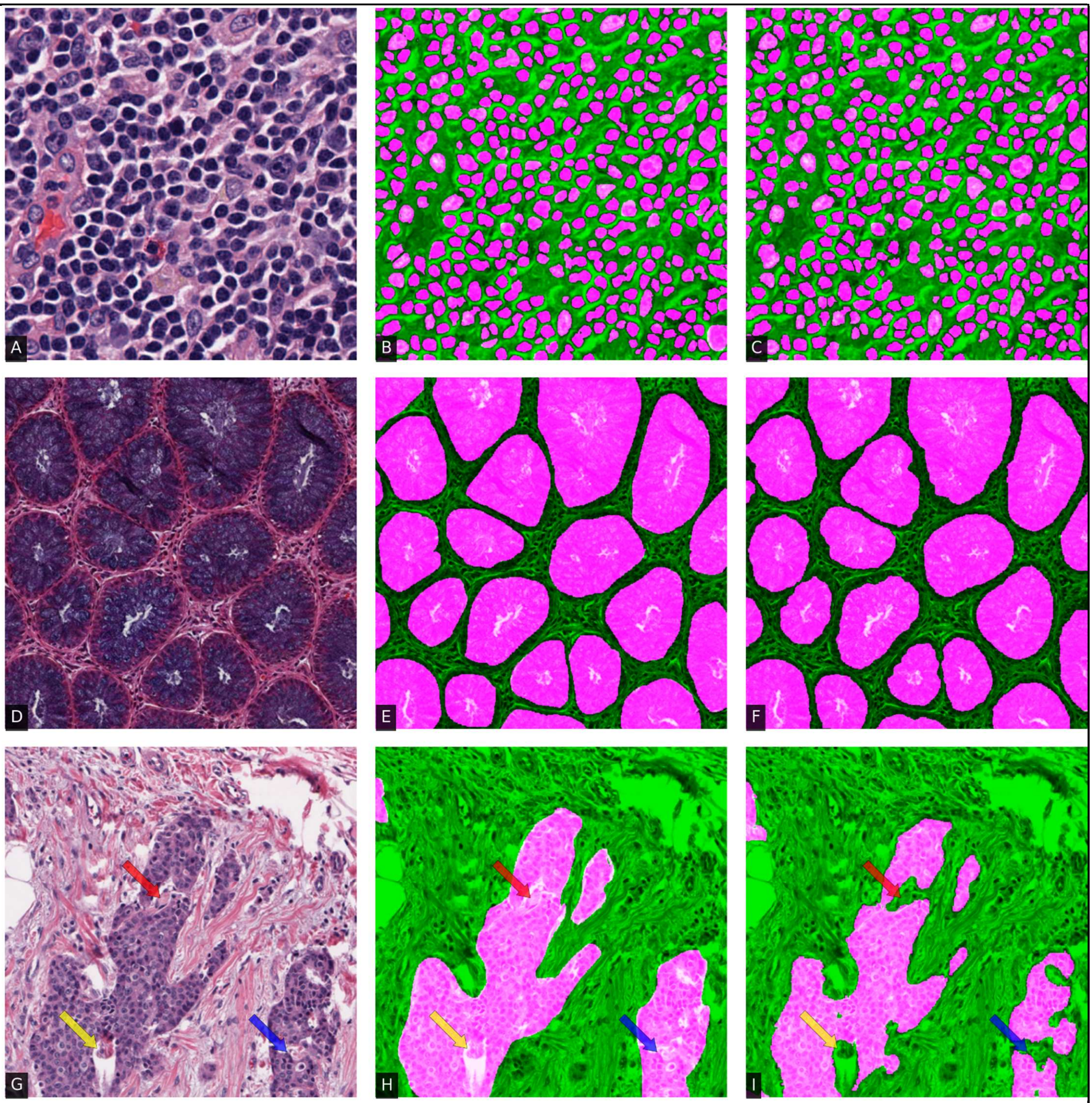

**Figure S1.** The figure shows original ROIs of (A) pancreatic nuclei, (D) colon tubule, and (G) breast cancers with associated (B & E & H) manual annotations and (C & F & I) QA annotation overlaid in fuchsia. The f-scores of QA versus manual annotation are (top) 0.97, (middle) 0.92, and (bottom) 0.91. In more complex regions (bottom row, areas

indicated with arrows), users often produce false positives, likely due to the associated additional time burden needed for intricate annotating.

Each use case followed the workflow presented in **Figure S3**. However, each use case benefited from slight variations in each step, due to histologic structure size, in order to optimize annotation efficiency (Section **SM 3**). All experiments were conducted on a Windows 10 desktop with a Nvidia RTX2060 8GB GPU.

| Histologic Structure | edgeweight [train_tl] | approxcellsize [superpixel] | compactness [superpixel] |
|---|---|---|---|
| **Cell Nuclei** | 8 | 20 | $10^{-4}$ |
| **Tubules** | 2 | 80 | $10^{-5}$ |
| **Epithelium** | 25 | 55 | $10^{-6}$ |

**Table S2.** Hyper-parameters (*edgeweight*, *approxcellsize*, and *compactness*) are set to different values for different structures.

**SM 2. Hyperparameters:**

QA is shipped mostly fully configured and the few hyperparameters of interest are easily modifiable via the configuration file (**Table S2**). In the use cases discussed here, the hyperparameters requiring modification were those governing the behavior of the superpixels on account of the target structure size. Briefly, a superpixel is defined as a group of adjacent pixels sharing similar characteristics in terms of chromatic, texture, or deep learned feature values[15] (**Figure S4**). The variable *approxcellsize* is set to the approximate width of the desired superpixel, and works well when set to the approximate width of the structure of interest. The nonnegative *compactness* value determines the regularity of the superpixel boundary, wherein higher compactness encourages superpixels to retain their initial square shape, while lower compactness allows for greater boundary irregularity. As an example, our epithelium use case employed a lower compactness setting due to highly irregular boundaries, versus nuclei which tend to be more consistently circular and thus have a smoother boundary. Lastly, setting a higher *edgeweight* encourages the DL model to focus the loss function on incorrectly classified boundary pixels; increasing this weight is beneficial when clear boundaries are hard to distinguish.

**SM 3. Use case specific workflows and insights:**

**SM 3.1 Nuclei Case:**

We selected 5 pancreatic cancer WSIs scanned at 40x from TCGA-PAAD dataset verified by Saltz's Group[21]. These 5 WSI images were divided into 2000 x 2000 image tiles. We selected 100 tiles from the generated ROIs. In accordance with the workflow presented in **Figure S3**, 20 nuclei tiles were uploaded into QA, a u-net autoencoder was trained, and patches were plotted on the embedding plot. Patches were then selected and manually annotated for 5 minutes. The DL prediction model was then trained and predictions were reviewed for modification and acceptance. The process iterated using batches of 20 tiles until all 100 tiles were completed.

In the first 5 minutes of nuclei annotation, even though no DL model was available, QA performed twice as fast as manual segmentation using QuPath[10] due to QA's superpixel functionality (.27 vs .14 nuclei per second). Superpixels enabled one-click selection for a subset of nuclei, notably improving annotation efficiency. As the DL model began to produce better predictions due to more training data, fewer modifications need to be made before accepting the model's proposals. This corresponded to the jump in improvement observed in **Figure 3A.**

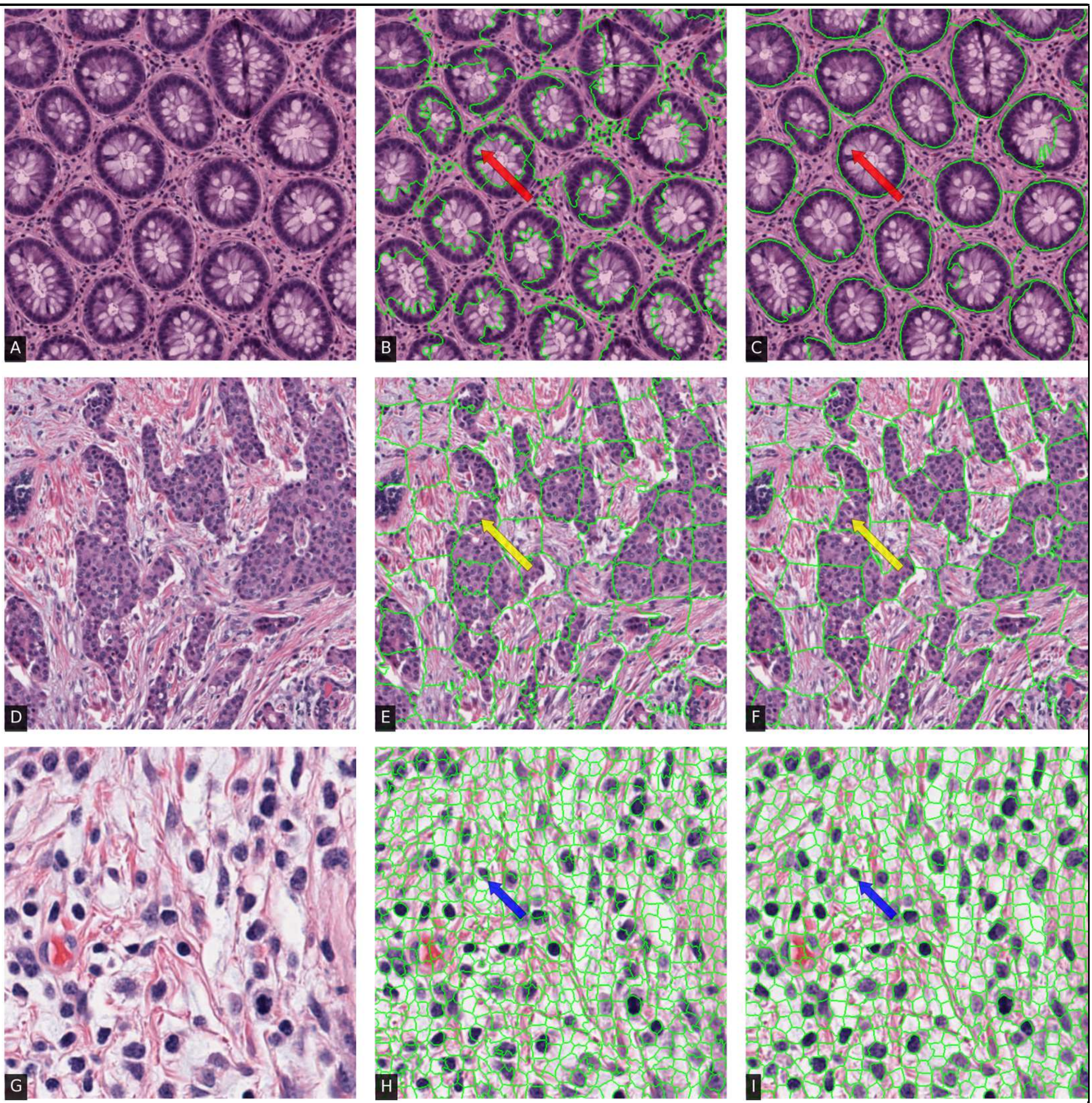

**Figure S4.** The (A & D & G) original 256 x 256 tubules, epithelium, and nuclei ROIs with (B & E & H) intensity-based superpixels and (C & F & I) deep learning derived superpixels. In QA, selecting a superpixel (boundaries shown in green) results in the groups of connected pixels sharing similar characteristics to be selected. As the user trains models, the superpixels evolve to capture more accurate boundaries of structures of interest. Later iterations (C & F & I) show increased specificity in terms of hugging structures of interest (e.g., areas indicated with arrows), thus enabling the user to rapidly select structures with high precision while using minimal effort.

**SM 3.2 Tubules Case:**

We selected 10 colorectal cancer WSIs from TCGA-COAD dataset. These 5 WSI images were divided into 1000 x 1000 image tiles and down sampled to 10x magnification, from which 100 tubule containing tiles were selected. To begin, 20 tiles were uploaded into QA after which the same workflow as to the nuclei use case was employed.

**Figure 3B** shows the efficiency changes over time as more tubules are annotated and the DL performance improves. Performance fluctuations were the result of differences in quality of WSIs, resulting in some tiles requiring additional correction. The superpixel feature continued to display a one-click selection for many tubules (**Figure S4**). Compared with nuclei annotation, tubule annotation efficiency converged faster and gave reliable suggestions with fewer annotated patches. The large difference in efficiency performance between nuclei and tubules (103x vs 9x, respectively) resulted from the fact that tubules occupy larger area, and thus there are fewer of them per 1000 x 1000 tile, implying more time is spent transitioning between tiles.

**SM 3.3 Epithelium Case:**

We selected 10 WSIs from an in-house estrogen receptor positive (ER+) breast cancer dataset scanned at 40x, and were processed similar to the tubule use case.

In the first 5 minutes of epithelium annotation, the bulk of the effort was spent manually delineating regions, as superpixel boundaries were not reliable (**Figure S4 yellow arrow**). This manual process is observed to be slower than the other 2 use cases due to the epithelial compartment's intricate structure. It appears that once a sufficient training set is created, coinciding with 246 annotated regions, the user starts to largely accept the DL suggestions. After this transition point, QA starts to provide improvements in both efficiency and annotation precision. For example, QA was able to provide better pixel-level segmentations in delicate regions which may be intractable for manual annotators (**Figure S5**).

**SM 4 Supplementary References:**

21. Hou, L. et al. Dataset of segmented nuclei in hematoxylin and eosin stained histopathology images of ten cancer types. Scientific Data 7, 185 (2020).

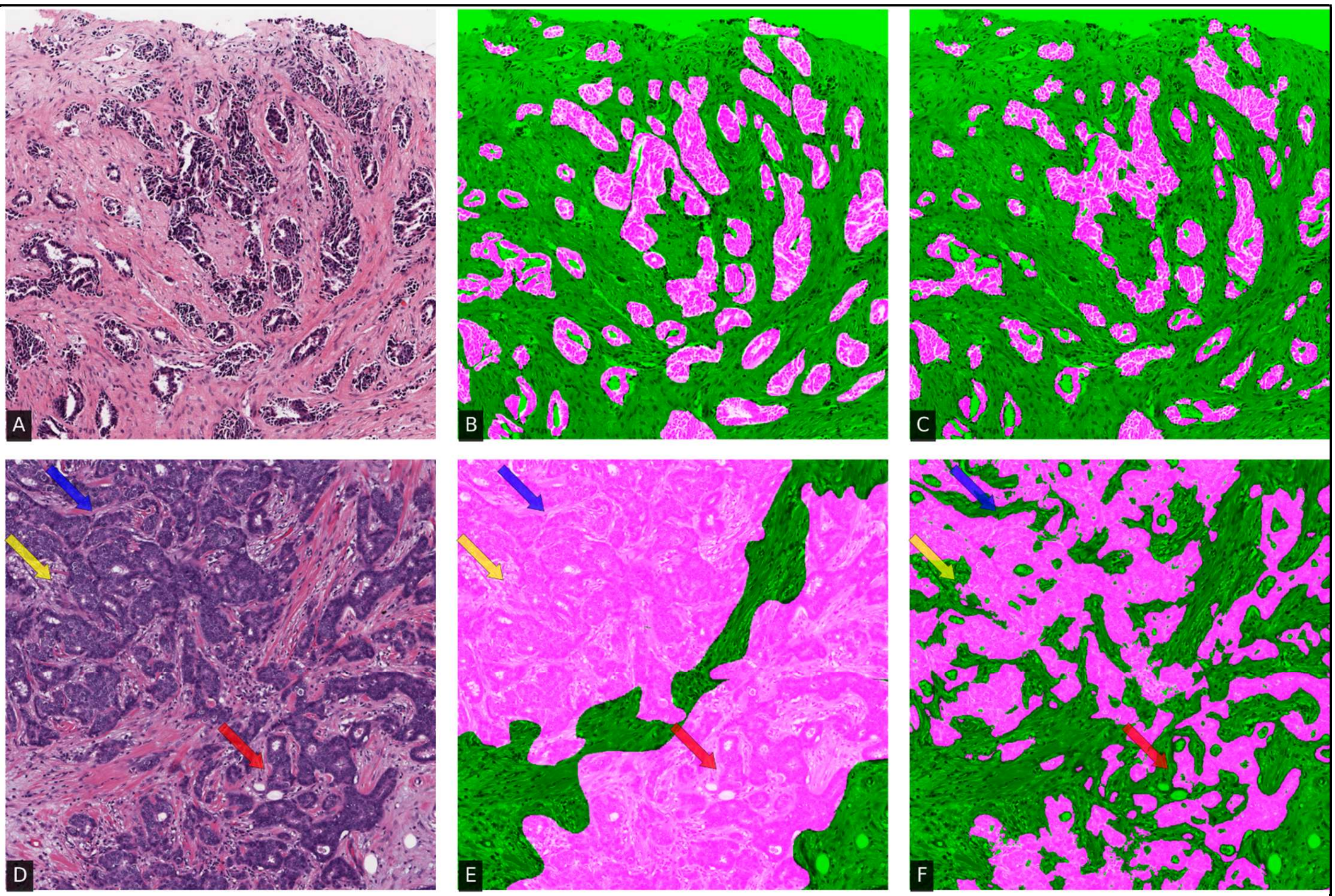

**Figure S5.** There is marked difference in levels of complexity between (A) less complex and (B) more complex epithelial regions. When manually annotating, large regions are more likely to be marked at a coarse level as indicated in fuchsia overlay (B & E). While QA is able to recapitulate the annotation with high fidelity in less complex images (C, f-score =.89), in more complex regions (F, e.g., areas indicated with arrows, f-score=.69) it appears to be able to provide a level of precision beyond that which would be achievable with human efforts.

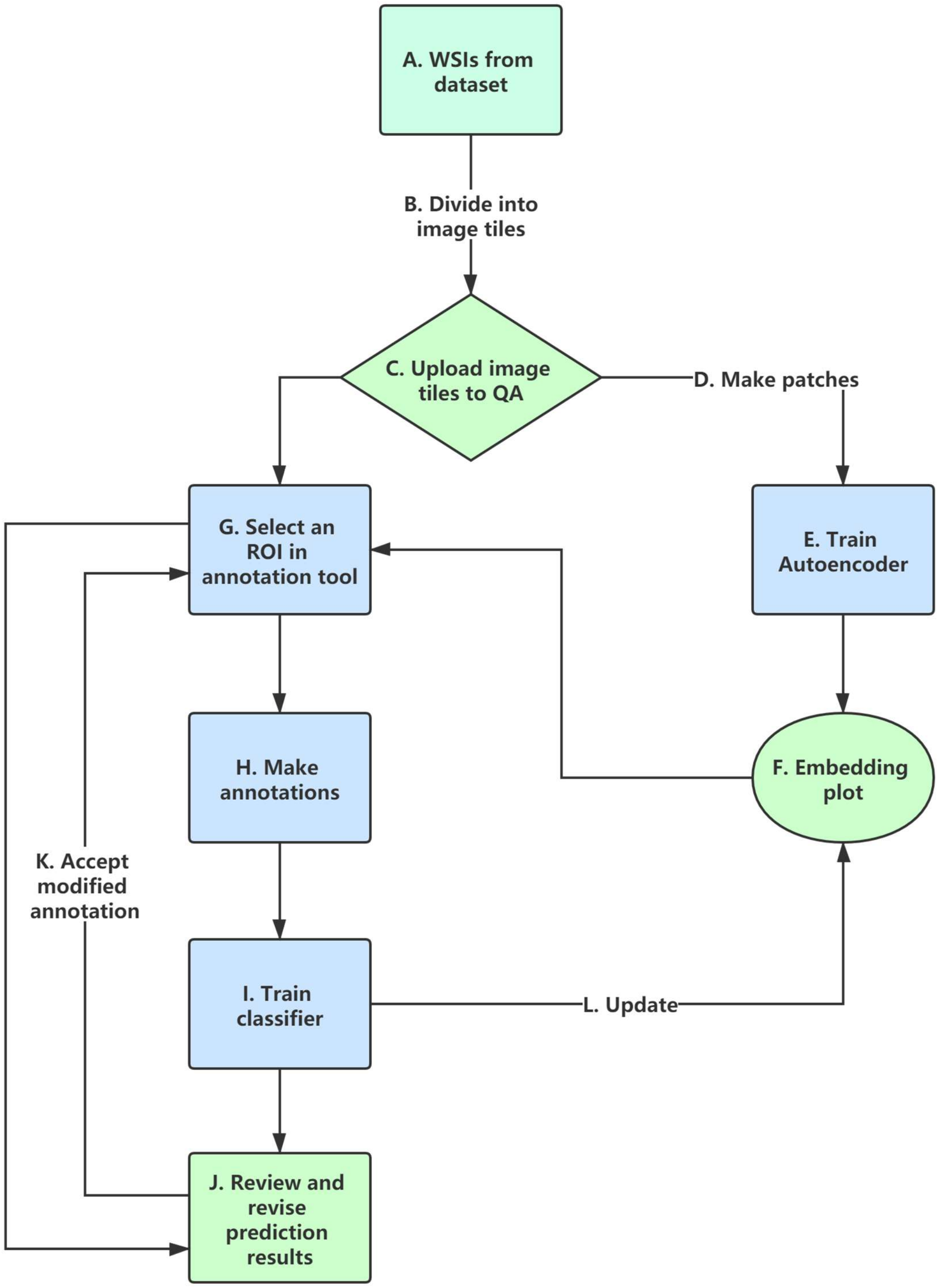

**Figure S3.** The flowchart illustrating general workflow of QA. (A & B & C) WSI are divided into tiles using a provided preprocessing script before being uploaded to QA. (D) Tiles are subsequently sub-divided into smaller patches of size 256 x256 amenable to deep learning. (E) These patches are used to train a u-net in an autoencoding fashion under the guise of initializing the u-net model with dataset-derived weights. (F) Each patch is subsequently embedded into a 2D space using UMAP from a feature vector derived from the transition layer between encoding and decoding of the u-net. (F -> G) The user then views the UMAP embedding plot (**Figure 1 A**) to select representative patches from the cohort for annotation. (H) Subsequently, the user annotates some suggested patches in the annotations pages (**Figure**

**1B**) with annotation aided tools. (I) These annotations were subsequently used to train the u-net and reapply it to the current image tile. (J & K) When viewing the annotation suggestions (**Figure 1C**), the user may either continue with manual annotation from scratch, or import the DL based suggestions for modification and acceptance of the annotations. (L) The embedding plot can be updated as the model improves, helping to identify poorly represented regions in the training set. As the model provides increasingly reliable suggestions, the user can begin to more confidently accept prediction results (**Figure 3**), allowing transitions directly from (G) to (J).